# A Meta-Heuristic Load Balancer for Cloud Computing Systems

Leszek Sliwko
University of Westminster, London, London, GB

Vladimir Getov
Faculty of Science and Technology, University of Westminster, London, United Kingdom

*Abstract*—This paper presents a strategy to allocate services on a Cloud system without overloading nodes and maintaining the system stability with minimum cost. We specify an abstract model of cloud resources utilization, including multiple types of resources as well as considerations for the service migration costs. A prototype meta-heuristic load balancer is demonstrated and experimental results are presented and discussed. We also propose a novel genetic algorithm, where population is seeded with the outputs of other meta-heuristic algorithms.

## I. INTRODUCTION

Modern day applications are often designed in such a way that they can simultaneously use resources from different computer environments. System components are not just properties of individual machines and in many respects they can be viewed as though they are deployed in a single application environment. Distributed computing differs from traditional computing in many ways. The sheer physical size of the system itself means that thousands of machines may be involved, millions of users may be served and billions of API calls or other requests may need to be processed [24][29].

In recent years the most advanced technologies offer cloud solutions [11]. A cloud system connects multiple individual servers and maintains the communication between them in order to process related tasks in several environments at the same time. Clouds are typically more cost-effective than single computers of comparable speed and usually enable applications to have higher availability than a single machine. This makes the software even more attractive as a service and is shaping the way software is built today [26]. Moreover, companies no longer need to be concerned about maintaining a huge infrastructure of thousands of servers just so they have enough computing power for those critical hours when their service is in highest demand. Instead companies can simply rent thousands of servers for a few hours [11]. Software as a service, where functionality is delivered to end users directly from distributed data centers, is a typical paradigm of the use of Cloud systems [26].

As of today, the biggest cloud systems offering elastic resource allocation are Amazon Web Services [22], Google App Engine [1], Microsoft Azure [27] Rackspace [27], Digital Ocean [24] and GoGrid [27]. A few well-known examples of services backed up by cloud computing are Dropbox, Gmail, Facebook, Youtube and Rapidshare. This elasticity of resources, without paying a premium for a large-scale usage, is unprecedented in the history of IT [11]. However, it introduces a new set of challenges and problems, which need to be solved. The cloud systems are usually made up of machines with very different hardware configurations and different capabilities. These systems can be rapidly provisioned as per the user's requirements [6] thus resource sharing is a necessity. Resource management has been an active research area for a considerable period of time and those systems usually feature specialized load balancing strategies.

This research outlines the significance of resource management strategies in cloud systems. This class of system is characterized by dynamic changes in their environments. The conventional container for applications in cloud are virtual machines, which can be quickly booted up or shut down on demand [10] and therefore the strategy needs to be robust enough to accommodate rapid changes in available resource configurations. To solve this problem, a number of strategies have been proposed and developed, with the majority focusing on achieving an optimal utilization of one resource exclusively [15][16][35].

In this paper, we specify an abstract model of cloud resources utilization (Section II), including multiple types of resources and consideration of service migration costs. Section III describes the research problem formulation. We then present the design of a prototype meta-heuristic load balancer (Section IV), which can be used to manage medium-size cloud systems. Sections V and VI provide the details of the experiments setup and results. In Section VII we conclude by discussing various employed strategies and highlight their advantages and weaknesses.

## II. MODEL OF CLOUD RESOURCES UTILIZATION

Our model consists of nodes and services where the load balancer task is to keep a good load balance through resource vector comparisons. In considering what is actually constituted as a 'service' in a Cloud environment an example may be seen in a popular Cloud environment such as Amazon's EC2, where applications are deployed within the full Operating System Virtual Machine (VM). Depending on design, service might come also with preinstalled local database such as MySQL. One might argue the effectiveness of this approach, however this schema has many benefits such as the almost complete separation of execution contexts (although services might still share the same hardware if they are deployed on the same node) and complete control over local system environment parameters. Amazon EC2 uses the templates in Amazon Virtual Image format [2]. Currently there are more than 20000 images to select from.

Services run constantly, which means they are not *tasks* which can be defined as a finite piece of work to be done [15] and do require resources, which are provided by the nodes. Every node has a certain amount of variable resource available, referred to in this paper as the *available resources* set. All resources on nodes are considered renewable and continuous, which means resources do not expire and cannot be depleted, assigning a service to node only temporarily lowers available resource levels. To simplify the definition, both the *resources needed* by the service and the *resources available* on the node are described by the vector of integer values. There exist several types of resources, which can be utilized by the service, such as memory, CPU cycles, and disk I/O operations. The number of defined resources is potentially unlimited, but in this experiment we use four types: CPU, memory, maximum network bandwidth and I/O operations speed. Services might have their resource needs shaped differently. There will be services experiencing hourly, daily or weekly variability in usage [30].

A Cloud system environment is characterized by very dynamic changes in resource availability. During its operations, some nodes might become idle or overloaded, additional resources might become available (new nodes might be added to network or demand for particular service decreases) or part of a cloud network could go offline. Therefore it is critical to provide a mechanism to automatically migrate services to alternative nodes.

Distributed systems often store or process large amounts of '*state*' - state consists of data such as database, files, relations, session data and identifiers, which are frequently updated [29]. Service migration is similar to jobs check-pointing [7]. During service migration the service Virtual Machine is stopped and its *state* saved to a state snapshot file. This file then gets copied over the network to an alternative node, where virtual machine is then restored. Therefore, a service will always carry some of the system *state* within itself. Saved state size depends upon how much data application has in memory and persistent storage. When we move a service to an alternative node, the *state* also has to be transferred. In this model, every service has its integer cost value assigned which is an abstract representation of the impact the *service migration* will have.

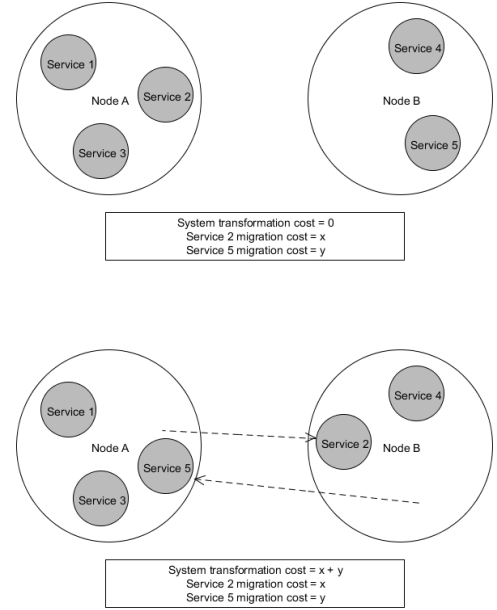

Figure 1. System transformation and migration cost

We consider the *service migration cost* value to be constant, as a migration of a certain service to any node will cause the same impact throughout the whole system.

## III. PROBLEM FORMULATION

Let us define $\Lambda = (\tau, \eta, \psi, a, r, c)$ as a problem space and system as a twice $(\Lambda, \mu)$. In the **d-resource system optimization problem**, we receive a set $\tau$ of $l$ mobile services $\tau = \{t_1, t_2, ..., t_l\}$ and a set $\eta$ of $m$ fixed nodes $\eta = \{n_1, n_2, ..., n_m\}$. We call $\mu : \tau \rightarrow \eta$ as a *service assignment* function, where each service has to be assigned to the node. We also consider:

- $\psi = \{i_1, i_2, ..., i_d\}$ as a set of all different kinds of resources. To illustrate, for $d = 3$ we could define $\psi = \{CPU, memory, network\}$.
- $a : \psi \times \eta \rightarrow \mathbb{N} \cup \{0\}$ as a fixed *available resources* on the nodes. $a_i(n)$ is the available level (integer value) of a resource $i$ on the node $n$.
- $r : \psi \times \tau \rightarrow \mathbb{N} \cup \{0\}$ as a fixed *required resources* for services. $r_i(t)$ is the required level (integer value) of a resource $i$ of service $t$.
- $c : \tau \rightarrow \mathbb{N} \cup \{0\}$ as a *service migration cost* function. $c(t)$ means cost incurred migrating service executables and its *state* and preparing service environment.

For every node $n \in \eta$ we define a set $A_n = \{t \in \tau : \mu(t) = n\}$ of all services assigned to the node $n$. We also define $f : \psi \times \eta \to \mathbb{N} \cup \{0\}$ as *remaining resources* on the nodes:

$$f_i(n) = a_i(n) - \sum_{t \in A_n} r_i(t) \quad (1)$$

We consider system $(\Lambda, \mu)$ as *stable*, if:
$f_i(n) \geq 0$, i.e.:

$$\sum_{t \in A_n} r_i(t) \leq a_i(n), \text{ for every } n \in \eta, i \in \psi \quad (2)$$

Otherwise the system $(\Lambda, \mu)$ is *overloaded*.

Each service $t$ is initially assigned by *service assignment* function $\mu_0$ to some node $\eta$. During the *system transformation* $(\mu_0 \to \mu_1)$ service $t \in \tau$ can be reassigned to any different node $n \in \eta$. The process of moving the service to a different node is referred to as *service migration* and this feature generates a *service reassigning cost*:

$$c_{(\mu_0 \to \mu_1)}(t) = \begin{cases} 0, & \mu_0(t) = \mu_1(t) \\ c(t), & \mu_0(t) \neq \mu_1(t) \end{cases}$$

Every *system transformation* process $(\mu_0 \to \mu_1)$ has its *system transformation cost*:

$$c_{(\mu_0 \to \mu_1)} = \sum_{t \in \tau} c_{(\mu_0 \to \mu_1)}(t) \quad (3)$$

Consider initial *service assignment* $\mu_0$; *service assignment* $\mu^*$ is optimal for $\mu_0$, if $\mu^*$ renders system $(\Lambda, \mu^*)$ *stable* and:

$c_{(\mu_0 \to \mu^*)} \leq c_{(\mu_0 \to \mu)}$, for every *stable* system $(\Lambda, \mu)$.

N.b.: when $(\Lambda, \mu_0)$ is *stable* for initial *service assignment* $\mu_0$, the *system transformation cost* equals $0$ as it is considered optimal.

We also consider two *service assignment* functions $\mu_0$ and $\mu_1$ to be *neighbors* if:

$$\left| \{t \in \tau : \mu_0(t) \neq \mu_1(t)\} \right| = 1 \quad (4)$$

The **d-resource system optimization problem** (D-RSOP) is a variant of classical Resource-Constrained Project Scheduling Problem (RCPSP), thus D-RSOP also belongs to the NP-hard (Nondeterministic Polynomial-time hard) problems class. Since its advent, RCPSP has been examined numerous times by researchers and numerous solutions have been proposed, implemented and tested [3][4][5][8][19][23]. RCPSP is solvable by simple heuristics such as the H1m (heuristic procedure where each job is assigned a fixed continuous resource amount equal to 1/m) and HCRA (heuristic procedure for continuous resource allocation) algorithms [21], however the result quality is low. Exact methods have been explored, but either they have a limitation of problem size or focus only on deriving new lower bounds as optimal solution can be found and verified only in small problem instances [19][28].

IV. LOAD BALANCER DESIGN

The general approach to solve job scheduling problems is to employ meta-heuristic strategies [19]. One can argue that meta-heuristics might be non-acceptable as solving load balancing problem in that each scheduling event can be very time consuming and have high overheads [25]. However, the resource management in a distributed system is nowhere near the dynamic and robustness level required on scheduling processes on CPU cores.

Load balancer prototype was implemented in a functional programming language Scala (version: 2.11.2). The source code of this load balancer is available at: (reference removed for review purposes). All computations were performed on a MacBook Pro with the following specifications:

| Model | MacBook Pro11,1 |
| --- | --- |
| Operating System | OS X 10.9.4 |
| CPU | 2.4GHz dual-core Intel Core i5 |
| Memory | 8GB 1600MHz memory |
| Storage | 256GB PCIe-based flash storage |
| Java Virtual Machine | 1.6.0_65-b14 Oracle |

Table 1. Experiment data – System configuration

The load balancer sequence was designed as follow:

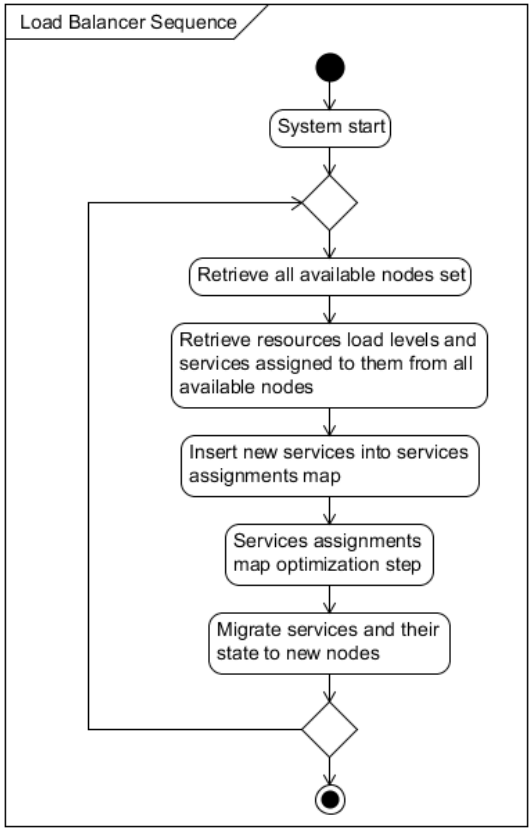

Figure 2. Load balancer sequence

The core of load balancer is a decision-making module, which assigns services to nodes. Every decision has computation overhead, yet badly assigned services can cause global system instability. Therefore, load balancer has to maintain a difficult balance between the speed and quality of its decisions. Selection of the most efficient algorithm is critical. Based on previous research [20], as well as our existing work not every algorithm will perform well with this problem. For the purpose of the experiment, we have selected several of the most promising strategies:

A. Greedy is an algorithm that follows the problem solving heuristic of making the locally optimal choice at each stage with the hope of finding a global optimum. In many problems, a greedy strategy is effective; however, it usually does not produce an optimal solution in this research. Nevertheless, a greedy heuristic will yield locally optimal solutions in a very quick time.

B. Tabu Search (TS) was introduced by Fred W. Glover in 1986 [13] and further formalized in 1989 [14]. This algorithm has been suggested by previous research on a similar problem [18]. TS searches for an improved solution in immediate neighbors (solutions that are similar except for one or two minor details). TS enhances its performance by maintaining a list of visited solutions so that the algorithm does not consider that possibility repeatedly.

C. Simulated Annealing is a general method for finding the global optimum by a process inspired from annealing in metallurgy (heating and controlled cooling of a material to increase the size of its crystals and reduce their defects [37]. This effect is implemented in the SA algorithm as a slow decrease in the probability of accepting worse solutions as it explores the solution space. Previous research over use of this strategy in load balancing can be found [20].

D. Genetic Algorithm (GA) — is a search heuristic that mimics the process of natural selection. GA belong to the larger class of evolutionary algorithms, which generate solutions to optimization problems using techniques inspired by natural evolution, such as inheritance, mutation, selection, and crossover. Unmodified GA has been previously examined with good results [19]. In this research we have deployed a variant with Genetic Drift step — detailed in [33].

E. Seeded Genetic Algorithm (SDA) — the generation of random solutions is the most costly step in GA strategy, sometimes taking up to 60-70% of a total computation time. Therefore, we have implemented a novel approach, where Genetic Drift step has been replaced with locally optimal solutions (i.e.: solutions seeding) found by Greedy, Tabu Search (TS) and Simulated Annealing (SA) algorithms. This approach should allow us to dramatically lower the total size of population (as individual genotypes are of better quality). Therefore, to test this approach, three respective strategy variations were created (SGA-Greedy, SGA-TS and SGA-SA).

F. Full Scan (FS) – this strategy performs a full search over all available configurations. FS strategy is convergent meaning it is able to find the globally optimal solution in finite time, under appropriate modeling assumptions. Multiple optimization techniques have been implemented in this algorithm, including shaving and path cut [9] and largest-migration-cost-first.

V. EXPERIMENTS SETUP

The characteristics of Cloud workload in data center are not perfectly clear yet and as research shows these will significantly differ from traditional grid computing [10]. There exists only a limited number of publicly available cloud system workload traces and those are generally anonymous and stripped of useful details [31].

The research community is mostly relying on simulations and models to conduct their experiments. The quality of input data and its realistic nature is a very important factor as it has a direct impact on the accuracy of results. There are a number of papers examining and providing statistical analysis [10][17][31][32] of available workload traces and a number of studies focus on creating a reliable workload traces generator [12][36].

In this experiment we have generated system configuration based on the above research and also on our professional experience with working with Amazon EC2 cloud instances. We have selected a set of services and nodes and proposed configuration shall be valid for the purpose of this experiment (table 2 and 3).

| Node | Resource I (CPU) | Resource II (Memory) | Resource III (Network) | Resource IV (I/O speed) |
|---|---|---|---|---|
| A | 100 | 50 | 100 | 70 |
| B | 70 | 40 | 70 | 50 |
| C | 50 | 80 | 70 | 50 |
| D | 60 | 60 | 50 | 80 |
| E | 50 | 90 | 80 | 40 |
| F | 60 | 100 | 50 | 60 |
| G | 80 | 50 | 50 | 40 |
| H | 80 | 80 | 80 | 90 |

Table 2. Experiment data – nodes configuration

| Service | Initial node | Migration cost | Resource I (CPU) | Resource II (Memory) | Resource III (Network) | Resource IV (I/O speed) | Service | Initial node | Migration cost | Resource I (CPU) | Resource II (Memory) | Resource III (Network) | Resource IV (I/O speed) |
|---|---|---|---|---|---|---|---|---|---|---|---|---|---|
| 01 | A | 4 | 1 | 10 | 4 | 2 | 31 | J | 4 | 19 | 18 | 1 | 8 |
| 02 | C | 5 | 1 | 6 | 5 | 2 | 32 | H | 10 | 6 | 14 | 3 | 7 |
| 03 | G | 4 | 5 | 2 | 5 | 6 | 33 | I | 2 | 3 | 10 | 3 | 2 |
| 04 | A | 17 | 10 | 17 | 1 | 2 | 34 | E | 4 | 2 | 8 | 1 | 8 |
| 05 | D | 10 | 14 | 10 | 1 | 1 | 35 | E | 9 | 8 | 9 | 9 | 5 |
| 06 | C | 3 | 3 | 12 | 3 | 8 | 36 | F | 4 | 8 | 15 | 13 | 1 |
| 07 | C | 6 | 15 | 2 | 18 | 3 | 37 | G | 2 | 12 | 8 | 5 | 3 |
| 08 | F | 6 | 1 | 4 | 8 | 4 | 38 | E | 2 | 16 | 11 | 1 | 2 |
| 09 | D | 4 | 4 | 3 | 17 | 10 | 39 | D | 8 | 13 | 8 | 6 | 4 |
| 10 | D | 4 | 8 | 19 | 19 | 8 | 40 | G | 3 | 6 | 9 | 10 | 1 |
| 11 | B | 8 | 5 | 9 | 18 | 4 | 41 | I | 6 | 14 | 1 | 11 | 8 |
| 12 | I | 6 | 16 | 14 | 3 | 2 | 42 | H | 7 | 8 | 3 | 10 | 3 |
| 13 | G | 4 | 6 | 5 | 17 | 11 | 43 | F | 9 | 9 | 9 | 10 | 9 |
| 14 | E | 5 | 18 | 11 | 13 | 4 | 44 | A | 8 | 11 | 8 | 12 | 11 |
| 15 | F | 1 | 10 | 9 | 12 | 8 | 45 | C | 2 | 5 | 5 | 7 | 18 |
| 16 | A | 9 | 12 | 17 | 14 | 1 | 46 | G | 6 | 2 | 7 | 3 | 2 |
| 17 | D | 5 | 3 | 6 | 8 | 6 | 47 | J | 5 | 4 | 3 | 10 | 16 |
| 18 | B | 5 | 8 | 12 | 3 | 11 | 48 | H | 8 | 5 | 2 | 14 | 8 |
| 19 | C | 7 | 15 | 12 | 8 | 9 | 49 | B | 2 | 6 | 7 | 1 | 1 |
| 20 | G | 1 | 4 | 8 | 6 | 12 | 50 | I | 1 | 1 | 9 | 6 | 13 |
| 21 | F | 7 | 12 | 10 | 5 | 1 | 51 | G | 4 | 4 | 11 | 9 | 6 |
| 22 | G | 2 | 3 | 16 | 16 | 2 | 52 | L | 3 | 7 | 2 | 7 | 5 |
| 23 | H | 5 | 6 | 19 | 1 | 4 | 53 | E | 12 | 6 | 6 | 10 | 12 |
| 24 | D | 3 | 16 | 11 | 2 | 3 | 54 | J | 10 | 3 | 9 | 8 | 10 |
| 25 | F | 4 | 14 | 8 | 15 | 9 | 55 | K | 8 | 5 | 5 | 4 | 8 |
| 26 | G | 10 | 4 | 15 | 7 | 8 | 56 | H | 7 | 6 | 3 | 5 | 7 |
| 27 | B | 2 | 20 | 19 | 5 | 2 | 57 | A | 3 | 8 | 12 | 2 | 6 |
| 28 | B | 8 | 16 | 2 | 3 | 5 | 58 | F | 1 | 12 | 17 | 1 | 9 |
| 29 | G | 6 | 16 | 10 | 3 | 1 | 59 | F | 6 | 10 | 8 | 6 | 14 |
| 30 | F | 5 | 1 | 1 | 3 | 10 | 60 | C | 5 | 9 | 2 | 3 | 8 |

Table 3. Experiment data – services configuration

VI. EXPERIMENTAL RESULTS

Three of strategies (Greedy, Tabu Search and Simulated Annealing) were designed with the end state (i.e. no more steps were possible). If a strategy finished before given time it was continuously re-run and the best result selected. Number of runs significantly varied per strategy, especially in the lower sizes of solution space (figure 3).

Each algorithm creates a number of candidate solutions during their run. Deciding if a candidate solution is stable

(i.e.: no nodes are overloaded) tends to be the most expensive step in computations, with around 50-70% of CPU time (depending on tested strategy) spent on validation of solution feasibility routines. As an optimization, implementations were caching newly created solutions, meaning the same tasks assignment setup is never tested twice for being stable as the result is retrieved from memory. In the chart below (figure 4) we have plotted the average number of unique candidate solutions created in each test scenario.

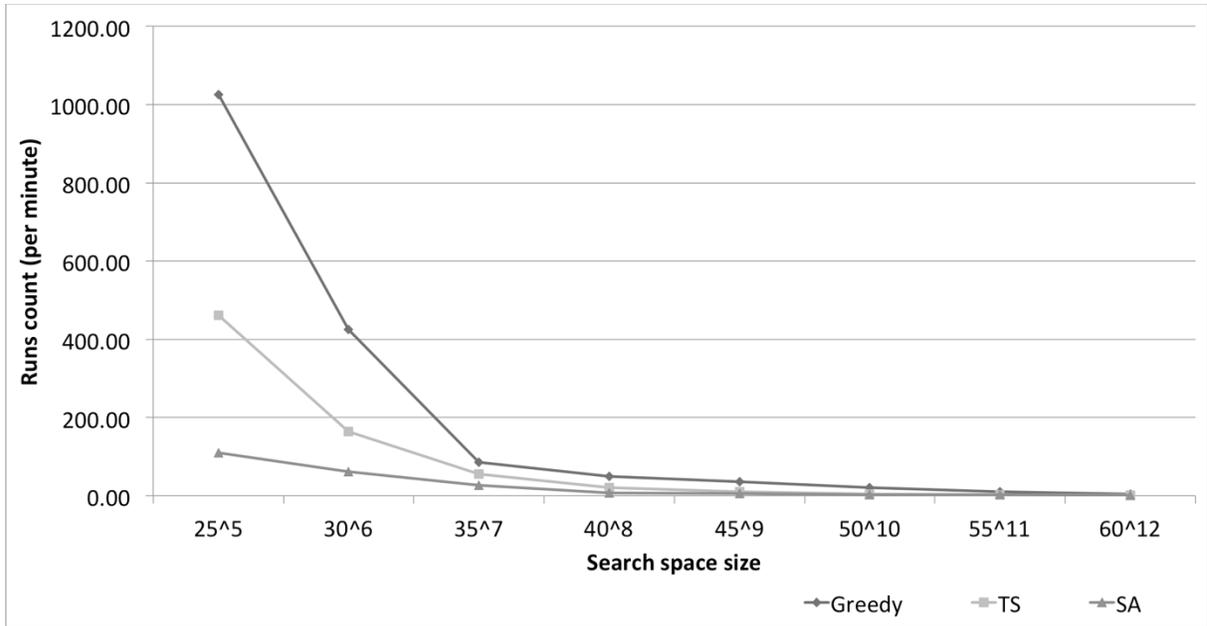

Figure 3. Runs count (per minute)

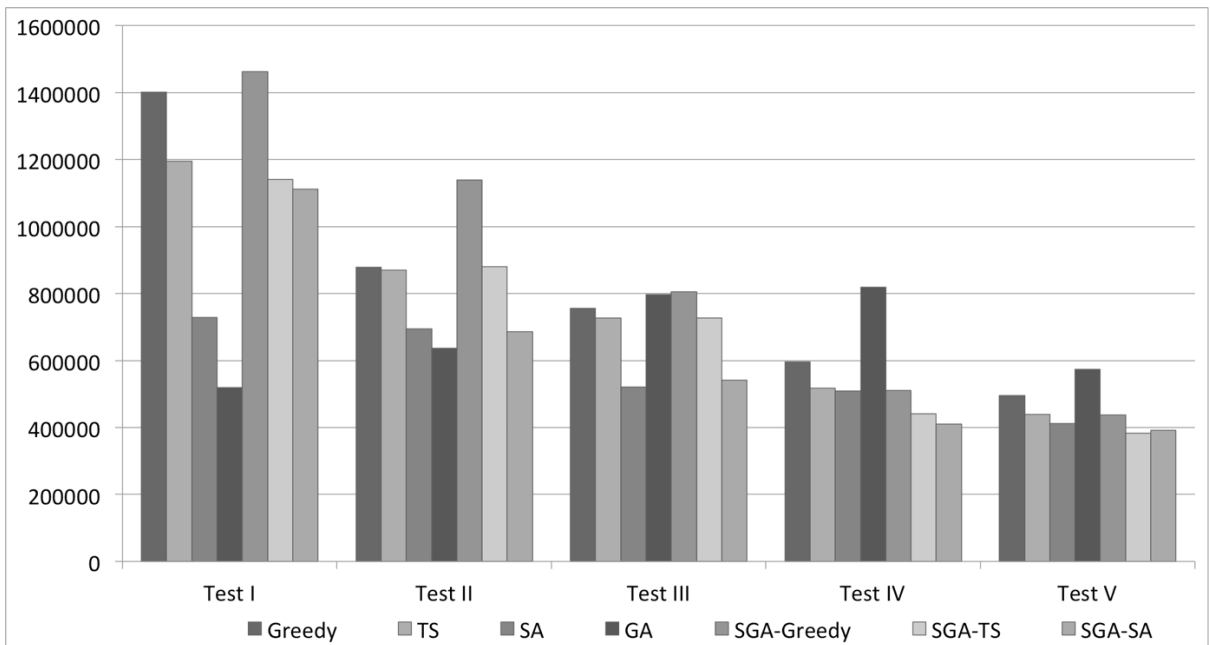

Figure 4. Unique candidate solutions created (per minute)

We designed five testing scenarios to see how each strategy copes with the increasing complexity of the problem. We have assumed new nodes are added only when new services are deployed [31] and a demand for computing resources increases. We are simulating this scenario with enabling additional nodes (in each test two additional nodes and ten more services are added). We have assumed that the load balancer will be run periodically, thus we have selected an arbitrary computation time, after which the best-found solution was selected as output result (table 4).

| Scenario | Deployed services | Enabled nodes | Computation time | Search space size |
|---|---|---|---|---|
| Test I | 1-20 | A-D | 30 seconds | $20^4$ |
| Test II | 1-30 | A-F | 1 minute | $30^6$ |
| Test III | 1-40 | A-H | 2 minutes | $40^8$ |
| Test IV | 1-50 | A-J | 4 minutes | $50^{10}$ |
| Test V | 1-60 | A-L | 8 minutes | $60^{12}$ |

Table 4. Experiment data – tests I, II, III, IV and V

The Full Scan strategy was used only as a benchmark if a global optimal solution was found and such limit was not imposed. The Full Scan strategy was not able to finish scenarios Test IV and Test V in reasonable time (24 hours and 5 days respectively). All other strategies' results were plotted on the chart above (the lower system transformation costs are better).

VII. RESULTS

As it was demonstrated in previous research 17][25][33], when solving classical Resource-Constrained Project Scheduling Problem and its variants, more complex meta-heuristics (e.g.: TS, SA, GA) perform significantly better than simple algorithms (e.g.: Greedy)

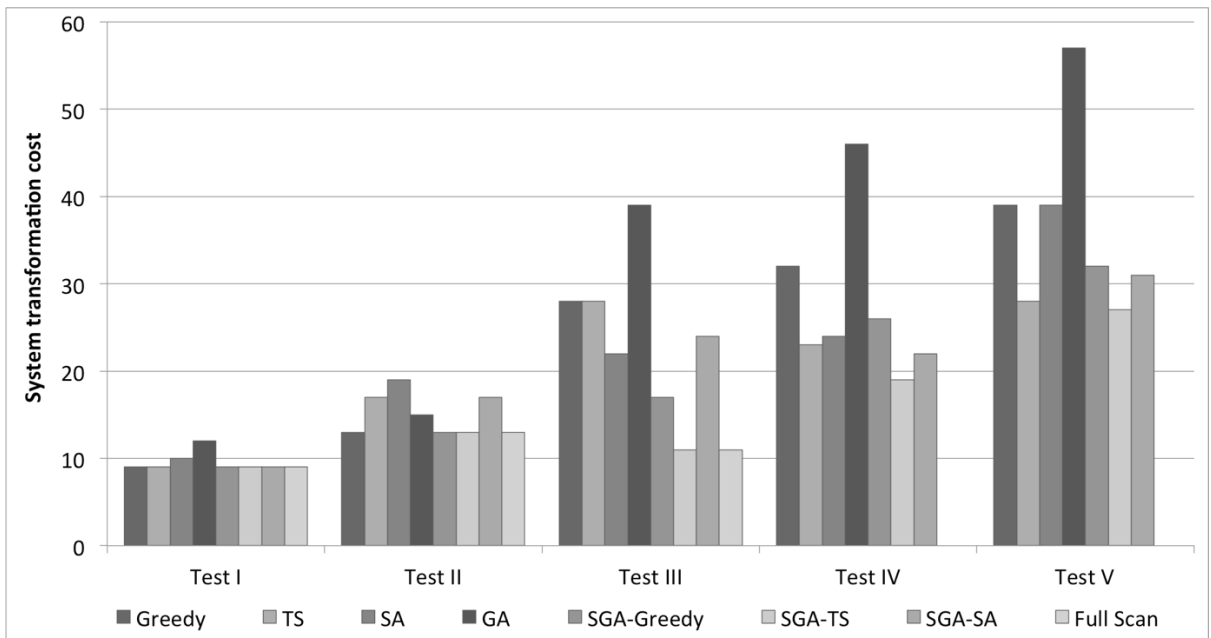

Figure 5. Results

A. Greedy – a very short execution time allowed strategy to be repeatedly run and therefore a few stable solutions were found in each test. Result solutions were of average quality; the most time consuming step was the generation of solution's neighbors (e.g.: during the Test V scenario, each step required 60 x 12 = 720 configurations to be examined).

B. Tabu Search (TS) – the main bottleneck in this approach was the last step where all of all same-value solutions had to be visited and marked as Tabu. Therefore, we have introduced a maximum limit of dull (without bettering solution) moves the strategy will perform, before the strategy gives up and returns the actual solution. Overall, the TS algorithm was working very well in small instances of a problem, which confirms results documented in [18].

C. Simulated Annealing (SA) strategy did require a much larger number of computations, often reaching only a fraction of runs in the same time as Greedy or TS. However, it did not require costly generation of all the solution neighbors, therefore re-runs count decreased at a much slower pace than above strategies. This strategy benefited the most from introducing the solution cache.

D. Genetic Algorithm (GA) variant was previously examined [33] and its main drawback is a costly generation of random solutions in the Genetic Drift step, especially when more types of resources are considered and a solution space grows in size. Performance was shown to be sufficient when examining two kinds of resources. However, due to the number of random generations required in order to create initial population the strategy performed quite poorly when four resources were introduced. As in [19], the larger problem size, the lower quality found solution was. This becomes noticeably apparent in instances of a larger problem, where ten or more nodes are involved.

E. Seeded Genetic Algorithm (SGA) was the most interesting strategy in our experiment. The introduction of solutions seeding to replace the previously designed Genetic Drift step [33] in the Genetic Algorithm allowed us to downsize the available genetic pool to 25% of its original size, which greatly reduced the computation time (around 50-70%) required to find good solutions without a reduction in quality. SGA returned the best results within the set time frame. In each case (Greedy vs. SGA-Greedy, TS vs. SGA-TS, SA vs. SGA-SA), the found solution was improved and generally less candidate solutions were examined (in Test V ca.14% less candidates were visited). In this experiment the variant with TS strategy returned the best results.

F. Full Scan strategy guarantees a globally optimum solution is found. Over the course of a research, this strategy has been heavily optimized: currently only about 9% of a solutions tree is traversed, the strategy starts moving services with the highest migration costs first, algorithm cuts leaves as soon as partial solution is deemed unstable. However, this still cannot be considered an efficient strategy due to a large number of computations required. In this experiment, Full Scan strategy was used to produce a global optima solution only in minor instances of a problem.

## VIII. CONCLUSIONS

In this paper after analyzing the algorithms performance, we came to the following conclusions, which might help us design new and/or enhance already existing algorithms:

1. The meta-heuristic algorithms rely on traversing a search space in small steps, meaning the next selected solution is usually similar to current one, but usually better. It might be beneficial to give higher priority to moving already-migrated services (as they already increased migration cost) and also prioritize moving services with smaller migration cost (due to reduced impact upon total migration cost). However, this step requires building problem-specific knowledge into algorithms.

2. The initial random generation of candidate solutions is expensive. This behavior is well visible in the upward trend in the number of candidate solutions created and tested (Figure 3) in Genetic Algorithms strategy. The number of tested solutions does not correlate with the quality of solutions and better results can be achieved if the solutions pool is initially created from already pre-computed set.

3. A few strategies succeed in reaching a certain solution level and they have difficulty to move out from this or recognize a last state (e.g.: only one neighbor solution is better). Especially the Tabu Search is prone to this issue. In this implementation we encountered a counter of steps without increasing quality of solution. When an arbitrary limit of steps is reached, strategy returns the current solution. However, we believe this can be handled in more intelligent way.

In our experiments, we tested the load balancer on a medium-sized networked system and found it capable of generating a huge ($60^{12}$ possible combinations) solution search space. We have also shown that increasing the number of tested resources did not hinder the performance of examined meta-heuristic strategies. In [33] we tested two resources, and in [34] we tested three resources. Finally in the current experiments a four-resource metric was used. Without doubt there is an opportunity to develop this area of research to focus on even more complex configurations.